\documentclass[aps,prd,twocolumn,showpacs,showkeys]{revtex4}
\begin{document} 
\title{A Lorentz covariant representation of bound state wave functions}  
\author{L.Micu}
\email{lmicu@theory.nipne.ro}
\affiliation{Department of
Theoretical Physics\\ Horia Hulubei Institute for Physics and Nuclear
Engineering, Bucharest POB MG-6, RO 077125}

\date{\today}

\begin{abstract}

We present a method enabling us to write in relativistic manner
the wave functions of some particular two particle bound state models in
quantum mechanics. The idea is to expand the bound state wave function
in terms of free states and to introduce the potential energy of the bound
system by means of the 4-momentum of an additional constituent, supposed
to represent in a  global way some hidden degrees of freedom.
The procedure is applied to the solutions of the Dirac
equation with confining potentials which are used to describe the
quark antiquark bound states representing a given meson state.

\end{abstract}

\pacs{11.10.St,12.39.Ki,12.39.Pn}

\keywords{bound states, quark models, relativistic covariance}

\maketitle

\section{Introduction}

About ten years ago a Lorentz covariant stationary expression for the
internal wave function of a meson has been introduced \cite{gm}. The meson
was supposed to be made of a free $q\bar{q}$ pair and of a collective
excitation of a background field representing the time averaged
result of the continuous series of quantum fluctuations taking
place in the bound system. In momentum space the generic form of the single
meson state was written as:
\begin{eqnarray}\label{meson}
&&\left.\vert {\mathcal M}(P)\right\rangle=
\int d^3k_1{m_1\over e_{1}}
d^3k_2{m_2\over e_{2}}d^4Q\nonumber\\
&&\times\delta^{(4)}(k_1+k_2+Q-P)\varphi(k_1,k_2;Q)\nonumber\\
&&\times\bar u_{s_2}(k_2)\Gamma_{\mathcal M} v_{s_1}(k_1)\Phi^\dagger(Q)~
\left.
b^\dagger_{s_1}(k_1)a^\dagger_{s_2}(k_2)
\vert 0 \right\rangle\,
\end{eqnarray}
where $a^+,~b^+$ are free creation operators of the valence $q\bar q$
pair, $u$ and $v$ are free Dirac spinors, $\Gamma_{\mathcal M}$ is a
Dirac
matrix  ensuring the relativistic coupling of the quark spins. The collective
excitation is represented by $\Phi^\dagger(Q)$ where $Q^\mu$ is the difference
between the bound state 4-momentum and the sum of the free quarks 4-momenta.

The ket (\ref{meson}) is not the output of a dynamical scheme. In fact, it is
only a phenomenological form with suitable features, like normalizability, 
Lorentz covariance, fulfillment of the mass shell constraints by the meson 
and the quark momenta.

Recently we found out that it is possible to derive an expression
like  (\ref{meson}) from the solutions of some bound state
problems in relativistic quantum mechanics which allow the independent 
treatment of the quarks while preserving the translational invariance
of the wave function. 
Observing that these requirements cannot be simultaneously fulfilled if 
the bound system
is made of a $q\bar{q}$ pair only, we supposed the existence of an
additional constituent which represents some hidden degrees of freedom
and show that the wave function can be written in a form like (\ref{meson}).

In the following we present a method enabling us to relate the relativistic
invariant function $\varphi(k_1,k_2;Q)$ in Eq.(\ref{meson}) with the 
bound state wave function
in quantum mechanics and the 4th component of
the momentum carried by the additional constituent, $Q^0$, with
the potential energy of the bound system. 

The second section is devoted to the particular models in
relativistic quantum mechanics for bound states which are adequate to  
our purpose. The bound state wave function factorizes into solutions
of the Dirac equation with confining potentials and a function describing 
the effect of the hidden degrees of freedom on the quark system.

By comparing these models with the bound state models
in field theory and establishing a correspondence between 
their elements having similar r\^oles we make the conjecture that the 
additional constituent represents in global way the quantum
fluctuations of the background field which generate the binding.

In the third section we present the general method leading to a
relativistic representation of the bound state wave function in the
specific cases presented above. The method is applied to the
analytical solutions of the Dirac equation with confining potentials
which are used to write the meson wave function.

In the last section we give some brief comments on the relation 
existing between various approaches to the bound state problem. Also, 
commenting upon the particular way of including the quantum fluctuations 
in a relativistic stationary representation of a bound system we conclude 
that our method is an alternative to the light cone formalism for low and 
intermediate energy.

\section{Bound state models}

As mentioned in the introduction, our way to ensure the quark independence 
without violating the invariance at translations of the bound state wave 
function is to assume the existence of an additional
constituent of the bound system. This is reflected 
by the appearance in the wave function of new variables representing the 
relative quarks coordinates with respect to the additional constituent.

Next we present two different ways of introducing the additional constituent 
into the quantum mechanical formalism.

In the first case the quarks are bound together by a confining potential
which depends on their relative coordinate. The additional constituent is
defined as a representative of some hidden degrees of freedom which induce
a random motion of the center of mass of the $q\bar{q}$ pair with respect to
the meson coordinates in the rest frame of the bound system.
The wave function $\Psi$ of the bound system can be written as:
\begin{equation}\label{Psi1}
\Psi(\vec{r}_1,\vec{r}_2,\vec{\rho})=\chi(\vec{r}_1-\vec{r}_2) ~
\phi(\vec{R}-\vec{\rho}_0)
\end{equation}
where  $\vec{R}=\eta\vec{r}_1+(1-\eta)\vec{r}_2$ is the position vector of 
the center of mass of the $q\bar{q}$ pair if $\eta={m_1\over m_1+m_2}$, 
$\vec{\rho}_0^i with i=1,2,3$ are the meson coordinates
and $\chi$ is the eigenfunction of the following Hamiltonian
\begin{eqnarray}\label{h0}
&&
{\mathcal H}=-i\vec{\nabla}^{(1)}\vec{\alpha}^{(1)}+
\beta^{(1)}m_1
-i\vec{\nabla}^{(2)}\vec{\alpha}^{(2)}+\beta^{(2)}m_2\nonumber\\
&&+ {\mathcal
V}^{(1,2)}(\vec{r}_1-\vec{r}_2),
\end{eqnarray}
whose eigenvalue is the meson mass, $M$.
In agreement with the above remarks, $\phi$ is the normal
distribution of the random variable  $\vec{r}-\vec{\rho}_0$
which writes as:
\begin{equation}\label{phi0}
\phi(\vec{r}-\vec{\rho}_0)={\mathrm e}^{-{(\vec{r}-\vec{\rho}_0)^2
\over 2\kappa^2}}
\end{equation}
where $\kappa^2$ is the variance.

In the second case, the additional constituent is associated with some
center of forces to which the quarks are independently
bound.  The wave function of the bound system reads:
\begin{eqnarray}\label{Psi2}
&&\Psi_M(\vec{r}_1,\vec{r}_2,\vec{\rho}_0)\nonumber\\
&&=\int~d^3\rho~
\bar{\psi}(\vec{r}_2-\vec{\rho})\Gamma_{\mathcal
M}\psi^c(\vec{r}_1-\vec{\rho})~ \phi(\vec{\rho}-\vec{\rho}_0),
\end{eqnarray}
where $\Gamma_{\mathcal M}$ is a Dirac matrix ensuring the relativistic
coupling of spins and $\psi$ is the eigenfunction of the single particle
Hamiltonian
\begin{equation}\label{h}
{\mathcal H}=-i\vec{\nabla}^{(j)}\vec{\alpha}^{(j)}+
\beta^{(j)}m_j+{\mathcal
V}_j(\vec{r}_j-\vec{\rho}).
\end{equation}
Here, ${\mathcal V}_j$ are confining potentials and the single particle
function $\psi^c$ defined as
$i\gamma_2\gamma_0\bar{\psi}^T$ is the charge conjugated partner of $\psi$.

Like in the first case the centers of forces are supposed to be randomly
distributed aroud the meson center, so that  $\phi$ writes as:
\begin{equation}\label{phi}
\phi(\vec{\rho}-\vec{\rho}_0)=
{\mathrm e}^{-{(\vec{\rho}-\vec{\rho}_0)^2\over2 \delta^2}}\,
\end{equation}
where $\delta$ is a free parameter.

We remark that in the above examples the additional
constituent does not behave like a real particle, but rather like an 
environment because it is not perceived through its specific properties,
but through its effect on the embedded quarks. This remark shall be enforced 
in the next section when making a comparison between the 
representations of the bound systems in
quantum mechanics and field theory.

\section{Lorentz covariant form of the wave functtion $\Psi_M$ }

The first step of the procedure leading to a relativistic expression
for the wave function it to expand  $\Psi$ in terms of free states whose
transformation properties at boosts are known.

Turning to the first case and noticing that
$\Psi(\vec{r}_1,\vec{r}_2,\vec{\rho}^0)$ in Eq.(\ref{Psi1}) is a
$4\times4$ matrix, we write:
\begin{eqnarray}\label{ps1}
&&\Psi(\vec{r}_1,\vec{r}_2,\vec{\rho})=\nonumber\\
&&\sum_{\epsilon_1,\epsilon_2,s_1,s_2}\int d^3k_1{m_1\over e_1}d^3k_2
{m_2\over e_2}{\rm e}^{i[\vec{k}\vec{r}+\vec{K}(\vec{R}-\vec{\rho})]}
\nonumber\\
&&\bar{\Psi}^{(\epsilon_1,\epsilon_2)}_{s_1,s_2}(\vec{k}_1,\vec{k}_2)
w_{\epsilon_1,s_1}
(\vec{k}_1)\bar{w}_{\epsilon_2,s_2}(\vec{k}_2)
\end{eqnarray}
where $w_{\epsilon,s}$ is the general notation for free Dirac spinors,
$\epsilon$ denotes the sign of the energy, $s$ is the spin projection on
an arbitrary axis. By $\vec{k}
=(1-\eta)\epsilon_1\vec{k}_1-\eta\epsilon_2\vec{k}_2$, $\vec{r}=
\vec{r}_1-\vec{r}_2$ and $\vec{K}=\epsilon_1
\vec{k}_1+\epsilon_2\vec{k}_2$ we denoted the relative momentum, relative
position vector and total momentum of the quark pair respectively.
According to the general principles of the quantum mechanics,
the coefficient $\tilde{\Psi}^{(\epsilon_1,\epsilon_2)}_{s_1,s_2}
(\vec{k}_1,\vec{k}_2)$ in (\ref{ps1}) writes as
\begin{eqnarray}\label{pm1}
&&\tilde{\Psi}^{(\epsilon_1,\epsilon_2)}_{s_1,s_2}(\vec{k}_1,\vec{k}_2)=
\int d^3r_1 d^3r_2 d^3\rho\nonumber\\
&&\times{\rm e}^{-i[\vec{k}\vec{r} +(\vec{R}-\vec{\rho})\vec{K}]} 
\bar{w}_{\epsilon_1,s_1}(k_1)\Psi(\vec{r}_1,\vec{r}_2,
\vec{\rho})~
w_{\epsilon_2,s_2}(\vec{k}_2)
\end{eqnarray}
and is the probability amplitude to find in the bound system a free 
$q\bar{q}$ pair with the
quantum numbers $\{\epsilon_1,\vec{k}_1,s_1\}$ and 
$\{\epsilon_2,\vec{k}_2,
s_2\}$ and a stationary wave with momentum
$-\epsilon_1\vec{k}_1-\epsilon_2\vec{k}_2$.

In the second case, using the same notations as above,
the single particle wave functions $\psi$ in (\ref{Psi2}), reads:
\begin{equation}\label{ps2}
\psi(\vec{r}-\vec{\rho})=\sum_{\epsilon,s}\int~d^3k~{m\over
e_k}~\tilde{\psi}^{(\epsilon)}_s(k)~w_{\epsilon,s}(k) {\rm e}
^{i\epsilon\vec{k}(\vec{r}-\vec{\rho})},
\end{equation}
where
\begin{equation}\label{pm2}
\tilde{\psi}^{(\epsilon)}_s(k)=\int~d^3r~\bar{w}_{\epsilon,s}(k)
~\psi(\vec{r})~{\rm
e}^{-i\epsilon\vec{k}\vec{r}}
\end{equation}
is the probability amplitude to find a free quark with positive (negative)
energy, spin $s$ and momentum  $\vec{k}$ in a bound state characterized
by $\psi$.
Taking into account the similar relations existing for $\bar\psi$ and for the
charge conjugated solution  $\psi^c$ we write the expansion coefficient 
of the wave function (\ref{Psi2}) as follows:
\begin{widetext}
\begin{equation}\label{tilde2}
\tilde{\Psi}^{(\epsilon_1,\epsilon_2)}_{s_1,s_2} (\vec{k}_1,\vec{k}_2)=
\int d^3\rho~{\mathrm e}^{i\vec{\rho}(\epsilon_1\vec{k}_1+\epsilon_2
\vec{k}_2)} \tilde{\bar{\psi}}^{(\epsilon_1)}_{s_1}(\vec{k}_1)
\Gamma_{\mathcal M}
\tilde{\psi}^{c(\epsilon_2)}_{s_2}(\vec{k}_2) \phi(\rho-\vec{\rho}_0).
\end{equation}
\end{widetext}

From the above examples it results that the additional
constituent contributes to the total momentum with $\vec{Q}=
-\epsilon_1\vec{k}_1-\epsilon_2\vec{k}_2$,which may be seen as the reaction
of the environement with respect to the independent motion of the 
quarks.

A similar relation is supposed to hold for the 4th components of the 
momenta where the ``energy'' of the additional constituent, $Q^0$, is
the binding energy of the system, defined
as the remaining part from the meson mass after extracting the
energies of the free quarks. Then we have:
\begin{equation}\label{Q}
Q^0=M-\epsilon_1\sqrt{\vec{k}_1^2+m_1^2}- \epsilon_2\sqrt{\vec{k}_2^2+
m_2^2}.
\end{equation}

In an ideal model the relation  (\ref{Q}), like the potential itself, should
result from the elementary processes, but for the time being this haven't 
been done. We notice however that nonrelativistic QCD resorts to a
definiton similar with (\ref{Q}) when introducing the effective potential
as the remaining part in the effective QCD Lagrangian after removing the
kinetic terms \cite{bb,pot}.

We remark that, according to the above definitions, 
the "square mass" of the effective constituent, $Q^2=
Q_0^2-\vec{Q}^2$, is not a constant and therefore, as we already
supposed, $Q^\mu$ is not the 4-momentum of an elementary particle.

Now,
introducing the occupation numbers of the free states and recalling
that in field theory the negative energy creation operators are
in fact annihilation operators of positive energy 
particles which give zero
when acting on the vacuum, we get the following expression for the ket
representing a single meson at rest:
\begin{widetext}
\begin{eqnarray}\label{meson1}
&&\left.\vert {\mathcal M}(M,\vec{0})\right\rangle=
\int d^3k_1~{m_1\over e_1}
d^3k_2~{m_2\over e_2}d^4Q
~\delta^{(3)}(\vec{k}_1+\vec{k}_2+\vec{Q})\delta(e_1+e_2+Q_0-M)
\nonumber\\
&&\times \sum_{s_1,s_2}\tilde{\Psi}^{(+,+)}_{s_1,s_2}(\vec{k}_1,
\vec{k}_2))
\bar{u}_{s_2}(\vec{k}_2)\Gamma_{\mathcal M} v_{s_1}(\vec{k}_1)
~\alpha^\dagger(Q)~ a_{s_2}^\dagger
(\vec{k}_2)~ b_{s_1}^\dagger (\vec{k}_1)\vert 0\rangle.
\end{eqnarray}
\end{widetext}
We mention that the quark (antiquark) creation
operators,
$a^\dagger$ ($b^\dagger$) are simple tools
reflecting the quark statistics, not the result of a canonical
quantification
formalism. Extending this notation to the additional constituent we
denote by
$\alpha^\dagger(Q)$ the creation of an effective excitation with
momentum
$Q^\mu$. The quark operators and $\alpha$ represent independent 
elements in a stationary representation and we find 
reasonable to assume they commute with each other. However, we notice that 
it doesn't make any sense to write commutations relations among 
$\alpha$-s because they do not represent elementary excitations and
hence they are not quantified. The best we can do is to  
deal with them in such a way as 
to ensure the overall conservation of the energy and momentum \cite{gm} in
the physical amplitudes which make use of expressions like (\ref{meson1}).

The last step towards a relativistic representation of
quark antiquark bound state is to write the expansion coefficients 
$\Psi^{(+,+)}_{(s_1,s_2)}$ in relativistic
invariant form. This can be easily done by replacing the 4th component of
a vector $V^0$  in the rest frame by $(P^\mu V'_\mu)~ M^{-1}$, where
$V'^\mu$ is the vector  $V^\mu$ in the reference frame where the meson
momentum is $P^\mu=(E,~\vec{P})$; also, the scalar products like
$\vec{k}_i \vec{k}_j$ must be written as $(P\cdot k'_1)(P\cdot k'_2)~M^{-2}-
(k'_1\cdot k'_2)$. 

Then turning to Eq.(\ref{meson}) we observe that it is equivalent 
with
(\ref{meson1}) if $\varphi(k_1,k_2;Q)$ is identified with
$ \tilde{\Psi}^{(+,+)}_{s_1,s_2}(k_{1},k_{2})$ in the first case and with
$\tilde{\psi}_{s_1}^{(+)}(k_{1})
\tilde{\psi}_{s_2}^{(+)}(k_{2})\tilde{\phi}(Q)$ in the second case, 
written with the Lorentz invariant notations defined previously.

\section{Examples}
\vskip0.5cm
{\bf The single particle Dirac equation}
\vskip0.3cm
The method presented in the previous section is now applied to the cases
where
the single particle wave functions $\psi$ are simple, analytical
solutions of the Dirac equation with confining potentials of the type
${\mathcal V}=\left(\begin{array}{c c}
{\mathcal V}_{1,2}&0\\
0&-{\mathcal V}_{2,1}
\end{array}\right)$ \cite{micu}.

\vskip0.3cm
{\it i. Linear rising potentials}
\vskip0.3cm
First we consider the case of linear rising potentials where
\begin{eqnarray}
&&{\mathcal
V}_1(\vec{r})=\zeta~r,\label{v1}\\
&&{\mathcal
V}_2(\vec{r})=-2m+\sqrt{\zeta\over\xi}
\left(1+2\vec{\sigma}\cdot\vec{L}\right)
+{\xi\over r},\label{v1p}
\end{eqnarray}
where the parameters $\zeta$ and $\xi$ characterize the potential.

The simplest analytical solutions having the angular
momentum $J$, magnetic number $M_J$, and energy
$E_J=m+2\sqrt{\zeta\over\xi}J$ are written as
follows:
\begin{equation}\label{psif}
\psi^{\rho}_{JM_J}(\vec{r})=\left(\begin{array}{r}
r^{\rho-1}{\mathrm  e}^{-\sqrt{\zeta\xi}r}~{\mathcal
Y}^{M_J}_{(J-{1\over2}) J}\\
-i\sqrt{\zeta\over\xi}~r^{\rho}{\mathrm e}^{-\sqrt{\zeta\xi}r}~{\mathcal
Y}^{M_J}_{(J+{1\over2}) J} \end{array} \right),
\end{equation}
where ${\mathcal Y}_{lJ}^{MJ}$ are eigenfunctions of the total angular
momentum $J$, and $\rho=J+{1\over2}$.

Projecting $\psi$ on the free Dirac solutions,
the expression (\ref{psif}) takes the form (\ref{ps2}). Using the
relativistic notation mentioned in the preceding section we get in the
case $J=1/2,~M_J=r$:
\begin{eqnarray}\label{plr}
&&\tilde{\psi}^{(+)}_{sr}(k)=\Omega^{(+)}(k)~\bar{u}_s(k)u_r(k),\\
&&\tilde{\psi}^{(-)}_{sr}(k)=\Omega^{(-)}(k)~\bar{u}_s(k){(\gamma\cdot P)
\over M}v_r(k)
\end{eqnarray}
where $\gamma^\mu,~\mu=0,1,2,3$ are Dirac matrices, $M$ is the meson mass 
and $\Omega^{(\pm)}(k)$ is given by the invariant form:
\begin{eqnarray}
&&\Omega^{(\pm)}(k)=\sqrt{2m~\zeta\xi\over e(k)\mp m}\left[\sqrt{\zeta\over\xi}
{\sqrt{e(k)\mp m}\over e(k)\pm m}\Sigma(k)\right.\nonumber\\
&&\pm
\left(1\left.+\sqrt{\zeta\over\xi}{1\over
e(k)\pm m}\right)\Delta(k)\right]
\end{eqnarray}
where
\begin{equation}
\Sigma(k)=
{\zeta\xi-3k^2
\over\left(\zeta\xi+k^2\right)^3}\,
\end{equation}
and
\begin{equation}
\Delta(k)=
{k\over
\left(\zeta\xi+k^2\right)^2}
\end{equation}
with $k$ representing the Lorentz invariant expression 
$k=\sqrt{(P^\mu k_\mu)^2 M^{-2}-m^2}$ and $e(k)=\sqrt{k^2+m^2}$.
\vskip0.3cm

{\it ii. Oscillator potential}
\vskip0.3cm
In the case of the oscillator potential where
\begin{eqnarray}\label{v2}
&&{\mathcal V}_1=\lambda~\omega^2~r^2,\\
&&{\mathcal V}_2=-2m+\lambda-2\omega(1+\vec{\sigma}\cdot\vec{L}),
\end{eqnarray}
the simplest analytical solutions of the Dirac equation are:
\begin{equation}\label{psios}
\psi_{J}^M(\vec{r})=\left(
\begin{array}{r} r^{J-{1\over2}}{\mathrm
e}^{-{1\over2}\lambda\omega r^2}~{\mathcal Y}^M_{(J-{1\over2}) J}\\
-i\omega~r^{J+{1\over 2}}{\mathrm
e}^{-{1\over2}\lambda\omega r^2}~{\mathcal Y}^M_{(J+{1\over2}) J}
\end{array} \right).
\end{equation}
Proceeding as above, and using the same notations, in the case
$J=1/2,M=r$ we get:
\begin{eqnarray}\label{ospl}
&&\tilde{\psi}^{(+)}_{sr}(k)=\sqrt{
e(k)+m\over 2m}\nonumber\\
&&\times\left[1-{e(k)-m\over \omega}\right]{\mathrm
e}^{k^2\over2\lambda\omega}~\bar{u}_s(k)~u_r(k),
\\ \label{osmi}
&&\tilde{\psi}^{(-)}_{sr}(k)=\sqrt{e(k)
+m\over 2m}\nonumber\\
&&\left[{e(k)+m\over \omega}-1\right]{\mathrm
e}^{k^2\over 2\lambda\omega}~\bar v_s(k){(\gamma\cdot
P)\over M}u_r(k). \end{eqnarray}
In all the above cases it is possible to define the charge conjugated
solutions $\psi^c=i\gamma_2\gamma_0\bar{\psi}^T$. As a result
$u\to v$ and
$\psi^{(\pm)}\to\psi^{c(\pm)}$  in (\ref{plr}-\ref{osmi}).

Closing this section, we mention that the single particle solutions
of the Dirac equation with confining potentials also can be used
in the case where the confining potential depends on the relative
quark coordinates if ${\mathcal V}^{(1,2)}(\vec{r}_1-\vec{r}_2)$ can
be written as the limit of ${\mathcal V}^{(1)}(\vec{r}_1-\vec{\rho}_1)+
{\mathcal V}^{(2)}(\vec{r}_2-\vec{\rho}_2)$ when $\vec{\rho}_1\to
\vec{r}_2$ and $\vec{\rho}_2\to\vec{r}_1$. Then the two particle
problem separates into two independent single particle problems
and the coefficient $\tilde{\Psi}^{(+,+)}_{s_1,s_2}
(\vec{k}_1,\vec{k}_2)$ can be calculated like in the previous cases.

\section{Comments and conclusions}

Now we comment briefly upon the features
of the different relativistic approaches to the bound state problem 
in order to get a deeper understanding of the subject.

First we consider the approaches in
relativistic quantum mechanics 
and observe that their common features are the existence of an attractive
potential well and of stationary wave functions with finite norms in the
space of the relative coordinates. Such approaches are not really
relativistic because the interaction potential cannot be written in a 
boosted reference frame.

In the approaches to the bound state problem derived from the field theory, 
the interaction is the result of the continuous exchange of quanta \cite{bs}
between the constituents. In this case, the iterative solution of the
dynamical
equation is expressed in terms of a relativistic interaction kernel and of
free propagators. As a result the solution is Lorentz covariant but has a
fluctuating character because the bound state appears to 
be made of an
indefinite number of free constituents.

The two representations look quite different. However, 
from consistency
arguments one may suppose the existence of a well defined 
correspondence
between their elements having similar r\^oles in the bound
 system.
In this sense we think reasonable to suppose that the
stationary wave function in the quantum mechanical approach
is the
result of a time average over a finite time $T$ of the 
fluctuating solution
in the field theory. This means that if the observation
time is longer than $T$, the fluctuating set of free particles
in the field
approach appears as a small, stationary set made of free 
particles, and of an effective excitation of the background field 
which gives rise to the binding.

In this paper we have shown that a similar 
representation emerges from a particular set of models with
confining potentials, if one assumes that the bound system 
contains beside the pair of quarks an 
additional effective constituent. Owing to this one we 
succeeded to write the bound 
state wave function as a superposition of free 
states and to escape
the problems raised by the presence of the interaction potentials by 
giving a Lorentz covariant meaning to the potential energy of
the bound system. 
Furthermore, by relating the effective constituent
to the quantum fluctuations of the background field generating
the binding we provided a justification 
for the existence of 
some spacial degrees of freedom accompanying
the interaction potential. These ones, which are quite unusual in 
quantum mechanics, in our models are the natural consequence 
of the imperfect cancellation of the vector momenta during the quantum 
fluctuations. 

Also, as an unquantized element of the bound system, the
additional constituent 
creates the possibility to by-pass the
Dirac no-go theorem  \cite{pamd} which states that a
bound system with a fixed number
of particles can be quantized in relativistic manner only 
if the generators of the symmetry group depend explicitly on the 
interaction potential. 

Concluding these comments, one can say 
that the additional constituent makes 
our method an alternative to the light cone formalism for low 
and intermediate energy, where only the average effect of the quantum 
fluctuations can be observed.
 
Now, comparing the two particular models presented in the second 
section, we remark that
the first one is much like the usual potential
models, where the quarks are the sources of the forces which
bind them together. However, the influence of the 
environment which manifests through the random motion of the 
center of mass of the $q\bar{q}$ pair around some fixed point
is not to be neglected, because it is essential for the relativistic 
treatment of the bound system 
and, moreover, it may offer a solution to 
the old center of mass problem in relativistic quantum mechanics. 

In our second case the confining forces are mainly due to the 
environment which looks
like a glue in which both quarks are embedded.
This model is similar with the bound state models
where the wave function is factorized in terms of constituent wave
functions and the confinement is the
result of the independent interaction of the quarks with 
an effective
constituent like, for instance,
the bag in the bag models  \cite{bm}.

We further remark that in both these models, as an effect of the quark 
independence, the mass center of the quark
pair is not at rest in the rest frame of the meson. The center of mass 
is at rest in the limit   
$\kappa\to\infty$ and $\delta\to\infty$ when the functions $\tilde\phi$ 
simulate the $\delta^{(3)}(\vec{k}_1+\vec{k}_2)$ and the quarks lose their 
independence. In this case
the first model transforms into a classical two body one 
and
has a well defined nonrelativistic limit, while
the wave function of the second model writes as
\begin{eqnarray}\label{nr}
&&\Psi(\vec{r}_1,\vec{r}_2,\vec{\rho}_0)\to\nonumber\\
&&\int d^3k_1 {m_1\over e_1}~d^3k_2 {m_2\over e_2}
~dQ^0\delta^{(3)}(\vec{k}_1+\vec{k}_2)\nonumber\\
&&{\mathrm
e}^{i(\vec{k}_1-\vec{k}_2)(\vec{r}_1-\vec{r}_2)}
\tilde{\psi}^{c(+)}_{s_1}(\vec{k}_1)
\tilde{\psi}^{(+)}_{s_2}(\vec{k}_2)...,
\end{eqnarray}
and it may lack a suitable classical
correspondent.

A last comment concerns the place of the time in the
present formalism. Our models are stationary quantum 
mechanical models where time is a simple
parameter, not a real coordinate like in the field
approach. In the correspondence we established between 
the two representations of a bound system we conjectured that the
stationary wave function in quantum mechanics is the 
result of a time average
over a finite time $T$ of the fluctuating solution in the
field approach, where $T$ is a physical parameter of the model, 
not a coordinate.

This supposition makes the coordinate representation
inadequate for the relativistic treatment of a bound system and
explains why it is impossible to relate
the iterative field solution in the coordinate representation
with the stationary solution in the quantum mechanical 
approach, or to give a relativistic meaning to the last one.

In this paper we have shown that the adequate representation is the 
momentum one, under the condition to find a suitable way to take 
into account the 
stationary effect of the continuous series of quantum fluctuations
generating the binding.

\begin{acknowledgments}
The work was finished during author's visit at the 
Center of Theoretical Physics in Marseille in the frame of 
the Cooperation Agreement between CNRS and the Romanian 
Academy. The hospitality at the Center of Theoretical Physics 
is warmly acknowledged.
The author thanks Dr. Claude Bourrely for clarifying discussions 
and for the careful reading of the manuscript.

The financial support from the Ministry of Education and 
Research in the
frame of the CERES Programme under the Contract No.
125/2003 and from the
Romanian Academy through the Grant No.21/2004 is
gratefully acknowledged.
\end{acknowledgments}

\end{document}